\documentclass[
 aip,
 amsmath,amssymb,
 reprint,%
]{revtex4-2}
\usepackage[colorlinks=true,urlcolor=blue,citecolor=blue,linkcolor=blue]{hyperref}
\usepackage{graphicx}
\usepackage{dcolumn}
\usepackage{bm}

\usepackage[utf8]{inputenc}
\usepackage[T1]{fontenc}
\usepackage{mathptmx}
\usepackage{etoolbox}

\makeatletter
\def\@email#1#2{%
 \endgroup
 \patchcmd{\titleblock@produce}
  {\frontmatter@RRAPformat}
  {\frontmatter@RRAPformat{\produce@RRAP{*#1\href{mailto:#2}{#2}}}\frontmatter@RRAPformat}
  {}{}
}%
\makeatother
\usepackage[colorlinks=true]{hyperref}

\usepackage{orcidlink}
\makeatletter
\newcommand{\ORCID}[1]{{\orcidlink{#1}}}
\makeatother
\begin{document}

\preprint{AIP/123-QED}

\title[\href{https://doi.org/10.1063/5.0310791}{Comment on J. Appl. Phys. 138, 114402 (2025)}]{Comment on ``Determining angle of arrival of radio-frequency fields using subwavelength, amplitude-only measurements of standing waves in a Rydberg atom sensor'' [J. Appl. Phys. 138, 114402 (2025)]}
\author{Matthew Chilcott\ORCID{0000-0002-1664-6477}}
\affiliation{ 
Department of Physics, QSO-Quantum Science Otago, and Dodd-Walls Centre for Photonic and Quantum Technologies,
	University of Otago, Dunedin, New Zealand
}
\author{Niels Kj{\ae}rgaard\ORCID{0000-0002-7830-9468}}\email{niels.kjaergaard@otago.ac.nz}
\affiliation{ 
Department of Physics, QSO-Quantum Science Otago, and Dodd-Walls Centre for Photonic and Quantum Technologies,
	University of Otago, Dunedin, New Zealand
}%

\date{submitted 5 November 2025}

\begin{abstract}
\end{abstract}

\maketitle

\onecolumngrid
\begin{center}  
\end{center}
\vspace{-5em}
\begin{figure*}[b!]
    \centering
    \includegraphics[width=\linewidth]{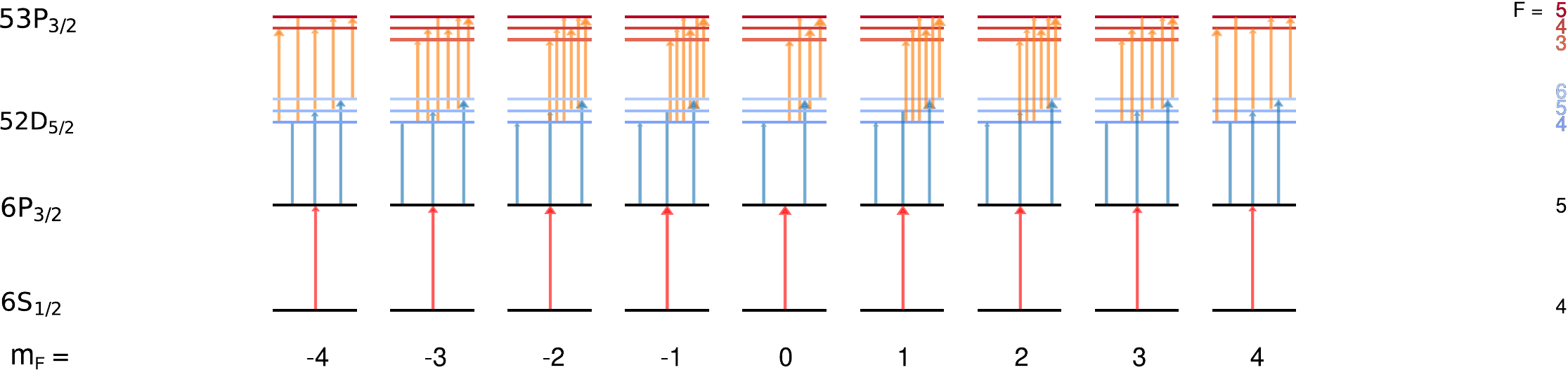}
    \caption{Reproduction of the truncated transition diagram Fig.~6(f) of Ref.~\onlinecite{Talashila2025} describing RF sensing via Rydberg EIT. The diagram omits the $F=1,2,3$ components of $52D_{5/2}$ Rydberg level and the $F=2$ component of the $53P_{3/2}$ Rydberg level. The $F$-components within each of the these two levels are energetically degenerate, but have been offset vertically to enable visual identification of transitions in play. Red, blue, and orange arrows represent, probe, coupling and RF field transitions, respectively (cf. main text). The area of the arrowheads is proportional to the transition strength.}
\label{fig:undressed_reduced}
\end{figure*}
\twocolumngrid
In a recent article, Talashila \textit{et al.}\cite{Talashila2025} describe the use of an optically-interrogated Cs vapour cell for determining the angle of arrival of an incoming RF field. The RF field dresses two atomic Rydberg levels which makes an imprint on the transmission of an optical probe beam propagating through the cell in a ladder scheme for electromagnetically-induced transparency (EIT) \cite{Finkelstein2023}.

In their Figure~6(f,g), the authors of Ref.~\onlinecite{Talashila2025}     present a diagram of atomic states of $\rm^{133}Cs$ relevant to the joint optical and RF interactions with the atoms of the vapour cell. It is stated that the diagrams are in the hyperfine basis and that they show the allowed transitions. We comment that neither of the two diagrams represent the physical situation. Specifically, the diagrams only depict 50 allowed RF transitions, where, as we shall argue, 84 RF transitions are in fact in play. This will affect the inferred optical response of the system.

\begin{figure*}
    \centering
\includegraphics[width=\linewidth]{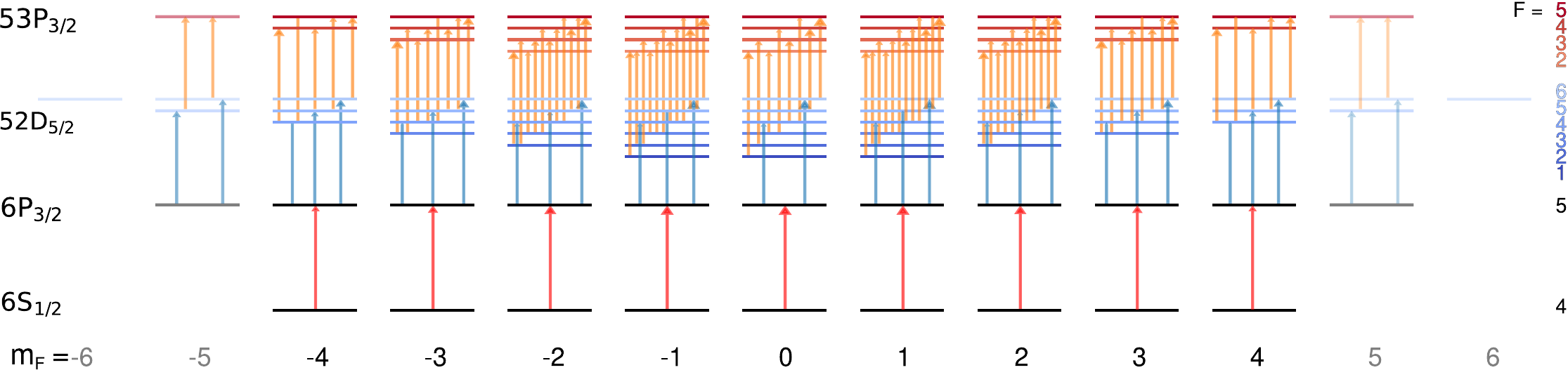}
    \caption{Extension of Fig.~\ref{fig:undressed_reduced} into an untruncated transition diagram. The states with $|m_F| \geq 5$ are shown faded to indicate that they not coupled to the measured probe field in the case where all the polarizations of all the fields are parallel.}
    \label{fig:undressed_full}
\end{figure*}

Figure~\ref{fig:undressed_reduced} reproduces Fig.~6(f) of Ref.~\onlinecite{Talashila2025}. The atomic ground level $6S_{1/2}(F=4)$ is linked to an intermediate level $6P_{3/2}(F=5)$ by a probe laser field, and the intermediate level is then linked to three energy degenerate Rydberg levels $52D_{5/2}(F=4,5,6)$ by a coupling laser field. This is the basis of ladder EIT \cite{Finkelstein2023}. An RF field then further couples each substate of the optically-addressed Rydberg level to substates of a higher-lying $53P_{3/2}$ Rydberg level. For example, $52D_{5/2}(F=4,m_F=1)$ is coupled to $53P_{3/2}(F=3,m_F=1)$, $53P_{3/2}(F=4,m_F=1)$, and $53P_{3/2}(F=5,m_F=1)$.
\begin{figure}
    \centering
\includegraphics[width=\linewidth]{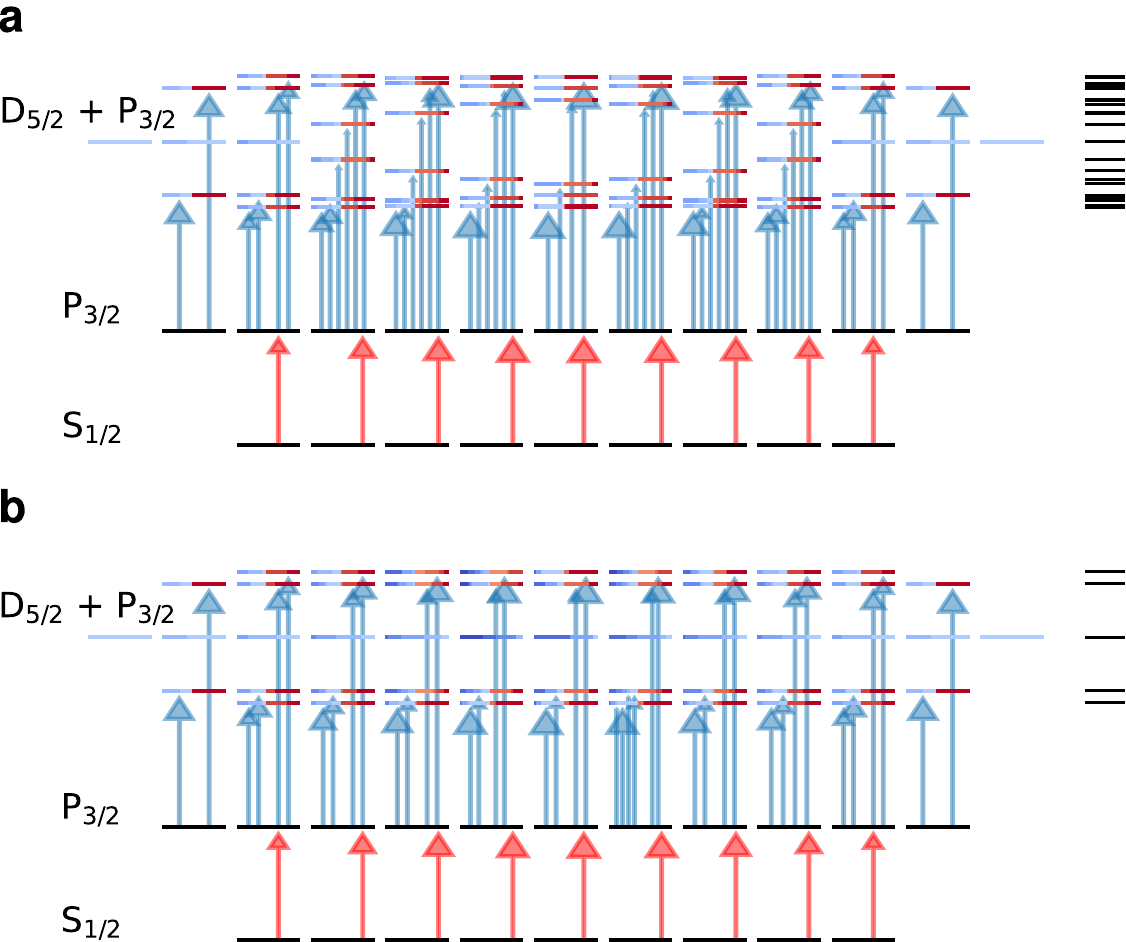}
    \caption{Optical transition diagrams for co-polarized fields with the RF-coupled Rydberg levels represented in a dressed state picture. Result of diagonalizing the RF coupling Hamiltonian for (a) the truncated system in Fig.~\ref{fig:undressed_reduced} and (b) full system in Fig.~\ref{fig:undressed_full}. The truncated and full system give rise to 25 and 5 unique eigenenergies, respectively, shown as projections on the right-hand side of the diagrams. The areas of the arrowheads represent the  transition strengths and we note that the central states are left uncoupled to the optical fields in both cases.}
    \label{fig:dressed}
\end{figure}.

The diagram in Fig.~\ref{fig:undressed_reduced} omits to show that the lower-lying Rydberg level contains hyperfine components $F=1,2,3$ and that the high-lying level includes an $F=2$ component. Despite there being no direct optical coupling to the $F=1,2,3$ components of the lower Rydberg level, these component cannot be excluded in the treatment: they are resonant linked by RF coupling to participating sub-states of the upper-lying Rydberg level. For example, the $P_{3/2}(F=3,m_F=1)$ sub-state of the upper level will couple to $D_{5/2}(F=3,m_F=1)$ and $D_{5/2}(F=2,m_F=1)$ of the lower level.  In turn, both of these will couple to $P_{3/2},F=2,m_F=1$ sub-state of the upper level, also absent in the diagram Fig.~\ref{fig:undressed_reduced}. We therefore submit that truncating the system is unwarranted and in the following we shall explore the consequence of leaving out states without direct optical coupling.

In Fig.~\ref{fig:undressed_full} we provide the transition diagram for the full system, when all fields, optical and RF, are co-polarized. While this diagram in the hyperfine basis appears more complex than that of Fig.~\ref{fig:undressed_reduced} and displays a significant increase in the number of RF transitions, the energy spectrum obtained for the Rydberg levels dressed by the RF field is, in fact, much simpler. Figure~\ref{fig:dressed} presents the truncated (Fig.~\ref{fig:dressed}a) and full (Fig.~\ref{fig:dressed}b) systems in a dressed-state picture. For the truncated system, the diagonalization of the RF coupling Hamiltonian reveals no less than 25 unique eigenenergies. Meanwhile, a field-dressing treatment of the full system gives rise to 5 unique eigenvalues in accordance with the rule given in Ref.~\onlinecite{Cloutman2024} applied to a $J=5/2\leftrightarrow 3/2$ transition.

For a linearly-polarized RF field, the number and values of eigenenergies obtained for the dressed Rydberg levels will not change when rotating the direction of polarization: the quantization axis can always be chosen along the RF field polarization. The innate energy spectrum of the system dressed in a linearly polarized RF field locks in the position of peaks in the optical transmission spectrum when the frequency of either the probe or the coupling field is scanned\cite{Cloutman2024}. The prominence of a peak in the spectrum can, however, change when the RF field polarization is rotated relative to the polarizations of the optical fields. For example, for $J=5/2\leftrightarrow 3/2$ transition, a central spectral peak is maximised for perpendicular polarizations and vanishes for co-polarized fields\cite{Sedlacek2013, Cloutman2025}. In addition, the relative prominence of spectral peaks will also depend on the field strengths involved.

\begin{figure*}
    \centering
    \includegraphics[width=\linewidth]{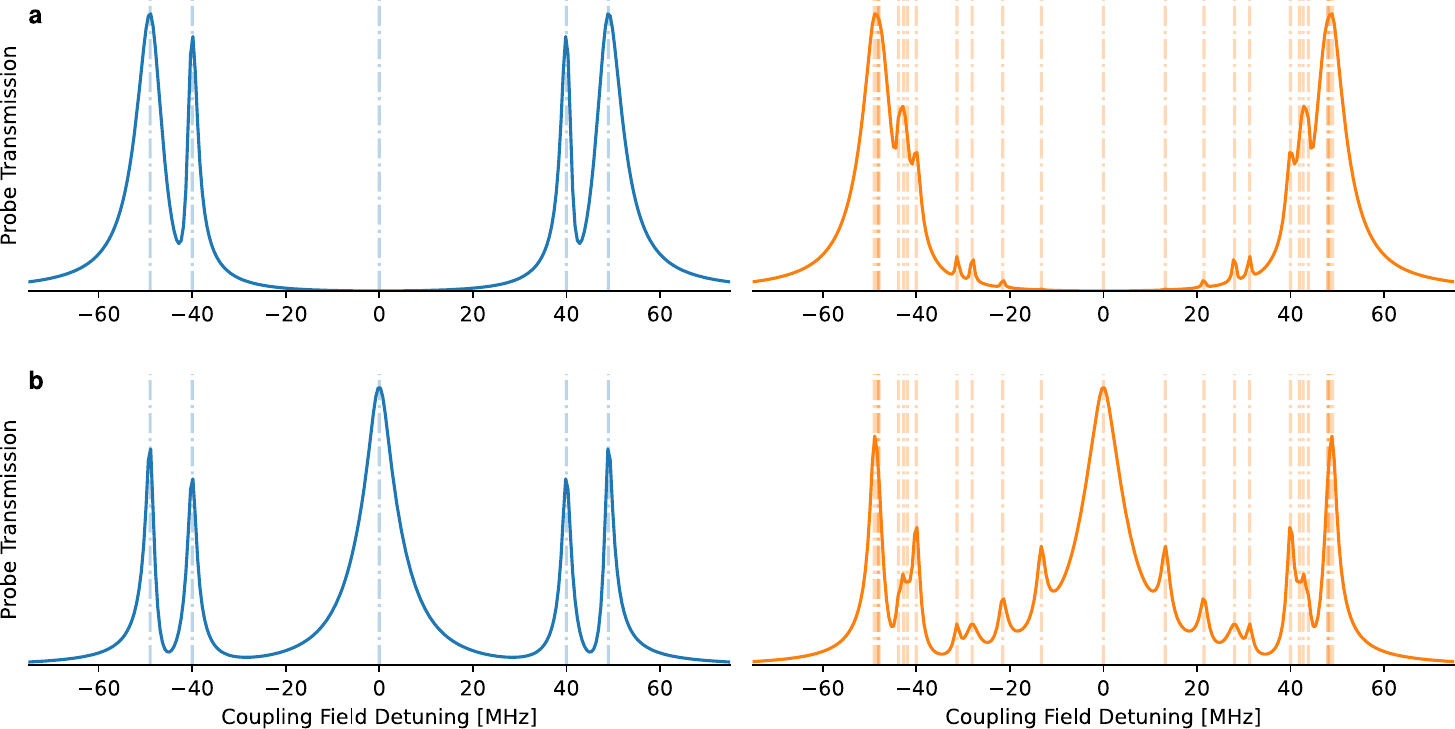}
    \caption{Simulated Doppler-free probe transmission when scanning the coupling laser frequency about nominal $6P_{3/2}\leftrightarrow52D_{5/2}$ resonance. Spectra for (a) co-polarized fields and (b) perpendicular RF-optical fields based on full (left-hand column) and truncated (right-hand column) transition diagrams are shown. Dashed vertical lines represent the energy eigenvalues of the RF-coupled Rydberg states. The simulations assume radial Rabi frequencies of 0.5, 20, and 200 MHz $ \times 2\pi$ for the probe, coupling and RF fields, respectively. }
    \label{fig:spectrum}
\end{figure*}

In Fig.~\ref{fig:spectrum}, we show simulated Doppler-free probe transmission spectra for a choice of field strengths corresponding to radial Rabi frequencies $\Omega_p=2\pi\times 0.5$~MHz, $\Omega_c=2\pi\times 20$~MHz, and $\Omega_{\rm RF}=2\pi\times 200$~MHz for the probe, coupling, and RF field, respectively. The transmission is calculated for nominally resonant probe light as a function of the coupling laser detuning. We present both the cases of co-polarized fields [Fig.~\ref{fig:spectrum}(a)] and perpendicularly polarized fields [Fig.~\ref{fig:spectrum}(b)]. The left-hand column shows the simple spectrum obtained for the full system corresponding to Fig.~\ref{fig:undressed_full}, while the right-hand column shows the more complex spectrum resulting from the truncated transition diagram of Fig.~\ref{fig:undressed_reduced}. As mentioned above, the details for the such spectra---for example, the prominence of particular features--- will depend on specifics such as field strengths. Also, the Doppler effect resulting from atomic movement and transit time broadening due to finite optical beam widths will act to smear out spectral features. But in general, when simulating the system, any prediction based on a truncated transition diagram will differ from a prediction based on a full transition diagram. Hence, our point stands that all the RF-transitions shown in Fig.~\ref{fig:undressed_full} should be included when accurately modelling the system.

\vspace{5mm}
\section*{Acknowledgments}
This work was supported by the Marsden Fund of New Zealand (Contract No. UOO2421).
\section*{Author Declarations}
\subsection*{Conflict of Interest}
The authors have no conflicts to disclose.
\subsection*{Author Contributions} NK wrote the original draft with input from MC. MC performed calculations and produced all figures. Both authors edited the final manuscript.
\section*{References}
\nocite{*}
\bibliography{comment}

\end{document}